\documentclass[onecolumn,showpacs,preprintnumbers,amsmath,amssymb,prb,superscriptaddress,nofootinbib]{revtex4}

\usepackage{amsmath}
\usepackage{amssymb}
\usepackage{graphicx}% Include figure files
\usepackage{epsfig}
\usepackage{dcolumn}% Align table columns on decimal point
\usepackage{textcomp}
\usepackage{bm}% bold math
\usepackage[normalem]{ulem} %need for strikethrough
\usepackage{color} %needed for coloring of \todo etc.

\newcommand{\ket}[1]{|#1\rangle}
\newcommand{\bra}[1]{\langle #1|}

\begin{document}

%\preprint{Technical notes in confidence - for internal use only - Version 1.10 (After review)}
\title{Scanning Quantum Decoherence Microscopy} 

\author{Jared H. Cole}
\address{%
Centre for Quantum Computer Technology, School of Physics, The University of Melbourne, Melbourne, Victoria 3010, Australia.
}
\address{%
Institut f\"ur Theoretische Festk\"orperphysik und DFG-Center for Functional Nanostructures (CFN), 
Universit\"at Karlsruhe, 76128 Karlsruhe, Germany
}
\author{Lloyd C. L. Hollenberg}
\address{%
Centre for Quantum Computer Technology, School of Physics, The University of Melbourne, Melbourne, Victoria 3010, Australia.
}

\date{\today}% It is always \today, today,

\begin{abstract}
The use of qubits as sensitive magnetometers has been studied theoretically and recent demonstrated experimentally.  In this paper we propose a generalisation of this concept, where a scanning two-state quantum system is used to probe the subtle effects of decoherence (as well as its surrounding electromagnetic environment).  Mapping both the Hamiltonian and decoherence properties of a qubit simultaneously, provides a unique image of the magnetic (or electric) field properties at the nanoscale.  The resulting images are sensitive to the temporal as well as spatial variation in the fields created by the sample.  As an example we theoretically study two applications of this technology; one from condensed matter physics, the other biophysics.  The individual components required to realise the simplest version of this device (characterisation and measurement of qubits, nanoscale positioning) have already been demonstrated experimentally.
\end{abstract}

\maketitle

The ability to image the very small is a powerful tool in the physical and life sciences.  For each advance in imaging techniques, there have been equally important steps in understanding the systems they image.  We theoretically show how a scanning two-state quantum system provides a unique perspective through monitoring of the subtle effects of decoherence.  Previous work has focused on using a coherent two-state probe as a sensitive spatial electrometer or magnetometer.  The system proposed here greatly expands this view by simultaneously imaging spatial information and the fluctuations occurring within.  To illustrate the technique, we investigate two contrasting applications - one from the realm of condensed matter physics, the other biological. Our results demonstrate that the technique provides a fundamentally new nano-imaging mode with wide applicability, combining both structure and process information in a way not accessible by other means and, given the recent demonstration of scanning qubit magnetometry, that implementation is well within reach of current technology. 

\section{Decoherence Microscopy}

Quantum coherence and entanglement in mesoscopic systems are such delicate effects that we usually treat the environment as an inevitable influence to be mitigated against at all costs. One seeks to isolate the quantum system as far as possible from the environment in order to maintain and observe non-classical effects. The imaging technique we introduce here is based on the inversion of this viewpoint - the inherently quantum properties of a probe system are deliberately exposed to an environment in a controlled fashion.  The effects of the environment are monitored and analyzed as a function of position. The resulting scan forms a unique image of the position and behavior of fluctuations in the environment as measured by the direct electromagnetic effects on the probe and its loss of quantum coherence.  

The use of qubits as sensitive magnetometers (or electrometers) has previously been discussed\cite{Chernobrod:05, Ilichev:07, Shnyrkov:07, DeSousa:06,Degen:08,Taylor:08} as the energy levels, and hence coherent evolution, depends very precisely on its electromagnetic environment.  Of particular interest to our proposal is the recent demonstration of a room temperature diamond colour-centre qubit scanning magnetometer system~\cite{Balasubramanian:08, Maze:08}.  Here we focus on a novel imaging mode, rather than increased sensitivity.  Monitoring the position dependent probe Hamiltonian and decoherence channels allows one to use the scanning quantum system as both a magnetometer/electrometer and as a probe of the decoherence environment \emph{simultaneously}.  As we will show, this imaging mode can reveal new features not observable by other means, as it is sensitive to the temporal as well as spatial variation in the fields created by the sample.  For a given sample, this provides more than just structural information, it also provides unique information on processes occurring in the sample and \emph{where} these processes occur.

The paper is organised as follows:  we first present the concept of scanning quantum decoherence microscopy in a non-specific way and illustrate its operation with a general example.  Following this, we present an example using existing technology, where we study a qubit composed of a nitrogen-vacancy defect within a diamond substrate.  We simulate the use of this system in imaging a biological macromolecule with non-trivial magnetic properties.

\begin{figure} [tb!]
\centering{\includegraphics[width=16cm]{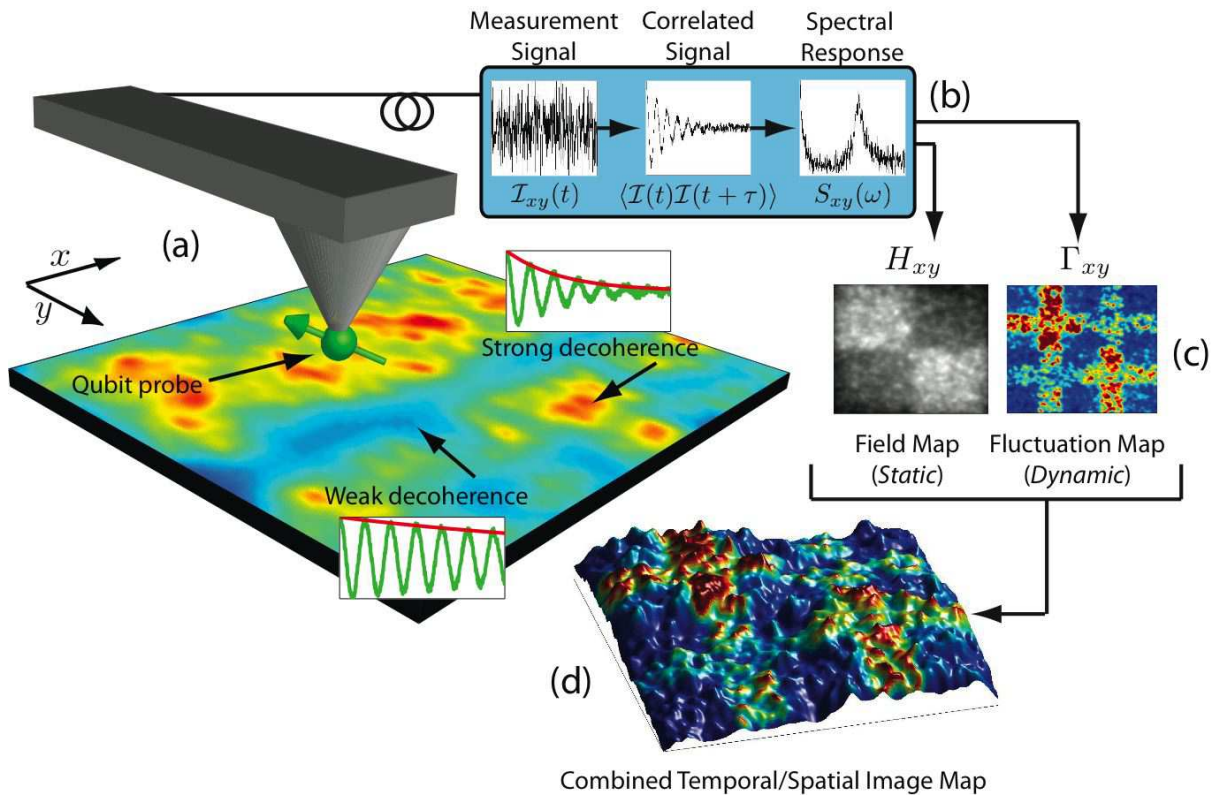}} 
\caption{Schematic of a scanning decoherence microscopy setup.  (a) The probe qubit is scanned across the sample while its quantum state is (weakly) monitored.  (b) At each point ($x$,$y$) a measurement record $\mathcal{I}_{xy}(t)$ is obtained.  Using the time-correlated signal, the spectral response $S_{xy}(\omega)$ of the qubit probe is determined. (c)  From this data, a measurement of the effective qubit Hamiltonian $H_{xy}$ (static magnetometer or electrometer) field map and decoherence rate $\Gamma_{xy}$ map as a function of probe position is obtained.  In this figure, the sample consists of fluctuators with a non uniform distribution in both spatial density and fluctuation rate.  The decoherence map reveals the fluctuator distribution not visible in the static field map.   (d)  Combining this information provides a direct window into the distribution and character of the sources of field fluctuations in both space and time.  In this image, the vertical scale is given by the strength of the field, whereas the colouration is given by the effective decoherence rate.\label{fig:schematic}}
\end{figure}%%%%%%%%%%

The imaging setup we envisage employs a two-state system (a qubit), based on charge or spin degrees of freedom, attached to the end of a cantilever probe, as illustrated in Fig.~\ref{fig:schematic}(a).  The quantum probe is governed by the Hamiltonian $H_p=\epsilon \sigma_x + \Delta \sigma_z$, where the $\sigma$'s are the Pauli operators and the coefficients ($\epsilon$ and $\Delta$) are functions of both the operating conditions and the probe's electromagnetic environment.  The probe is scanned across a sample and at each point, the Hamiltonian and decoherence felt by the qubit is determined through a series of measurements.  The decoherence of the qubit, in particular, is highly sensitive to the fluctuations in the local environment.  Thanks (in part) to the ongoing work to construct controllable quantum devices~\cite{Nielsen:00}, a number of techniques have been developed to measure the state or evolution characteristics of an (open) quantum system.  These include techniques to reconstruct an arbitrary quantum state or process\cite{Nielsen:00,Paris:04}, quantum channel\cite{Nielsen:00,Mohseni:08} and the Hamiltonian governing a few state quantum system\cite{Schirmer:04,Cole:05,Cole:06,Cole:06b,Devitt:06,Jordan:07} or even spin echo techniques~\cite{Poole:72} from magnetic resonance.  

The simplest realisation of decoherence microscopy is to perform a Rabi oscillation or spin echo experiment on a qubit which can be moved relative to the background decoherence source.  In this case, the measurement procedure (initialise, evolve, measure) is used to map the (ensemble averaged) time evolution of the system and the decoherence rate is fitted as an exponential decay.  With sophisticated measurement schemes and fitting procedures, the image can be acquired more quickly and efficiently.  In this paper, we will focus on the technique of Hamiltonian characterisation~\cite{Schirmer:04,Cole:05,Cole:06} and a general weak continuous measurement of the qubit as an example.  It is important to note that the particular choice of continuous measurement and Hamiltonian characterisation is not a requirement to perform decoherence microscopy, it simply provides an elegant and efficient realisation but other more established experimental techniques could be similarly applied.

We consider a quantum probe, located at position ($x$,$y$), suspended a distance $h_p$ above the sample.  The probe is weakly monitored, providing a measurement record $\mathcal{I}_{xy}(t)$.  This measurement record is then correlated and the spectral response $S_{xy}(\omega)$ computed, see Fig.~\ref{fig:schematic}(b).  The Hamiltonian and decoherence parameters can then be estimated from this spectral response, for each probe position.  If the components of the Hamiltonian are plotted as a function of position across the sample ($H_{xy}$), the probe acts as a sensitive electrometer or magnetometer (depending on the type of qubit), as given in the $H_{xy}$ example in Fig.~\ref{fig:schematic}(c).  However, complete analysis of the probe evolution allows the decoherence channel(s) to be extracted ($\Gamma_{xy}$), giving information about the strength, direction and character of the \emph{dynamics} of the environment, as well as the induced static field.  The resulting decoherence scan, shown in Fig.~\ref{fig:schematic}(c), reveals new information about the fluctuator frequency distribution in the sample which was not apparent in the electrometer/magnetometer image.  Combining both sets of data, Fig.~\ref{fig:schematic}(d), provides a direct image of the correlation between the temporal dynamics of the environment and the spatial structure.

To make our proposal quantitative, we introduce a model for weak continuous measurement~\cite{Brun:02} which captures all of the essential physics\footnote{While this is the more general scenario, for some systems strong projective measurement is a more natural implementation.  In this case, projective or `DC' measurements can be used to measure the system evolution~\cite{Schirmer:04,Cole:05,Cole:06}, producing equivalent results.}.  Here, we assume the measurement of the qubit can be modeled as a inefficient (or weak) positive operator valued measure (POVM) in the $\sigma_z$ basis.  The density matrix after measurement, $\rho'$, is given by
\begin{equation}\label{eq:POVM}
 \rho' = \frac{A_\pm \rho A_\pm^\dag}{\rm{Tr}[A_\pm^\dag A_\pm \rho]},
\end{equation}
for a measurement operator 
\begin{equation}
 A_\pm = \frac{1}{\sqrt{2}}\left(\sqrt{1\pm \kappa}\ket{0}\bra{0} + \sqrt{1 \mp \kappa}\ket{1}\bra{1}\right),
\end{equation}
with some measurement strength $\kappa$.  The measurement process consists of repeated weak POVM measurements separated by a time interval $\Delta t$, during which time the system undergoes normal evolution.  The measurement repetition interval is then a measure of the bandwidth of the detector, $\rm{BW}=1/\Delta t$.

The measurement record $\mathcal{I}(t)$, is the result, $+1$ or $-1$, of a measurement at time $t$.  The steady-state autocorrelation of this measurement signal $\mathcal{I}(t)$ is then given by
\begin{equation}
\langle \mathcal{I}(t)\mathcal{I}(t+\tau) \rangle_{\rm{ss}} = \rm{Tr}[\sigma_z e^{\mathcal{L} \tau} \sigma_z \rho_{\rm{ss}}]
\end{equation}
via the quantum regression theorem\cite{Gardiner:91, Scully:06}.  Here $\rho_{\rm{ss}} = \rho(\infty)$ is the steady state density matrix and $e^{\mathcal{L} \tau}$ is the solution to the density matrix evolution governed by $\dot{\rho}=\mathcal{L}[\rho, H, \Gamma]$, where $H$ is the qubit Hamiltonian and $\Gamma$ represents the decoherence rate of the system without measurement.  The spectrum of the signal is then
\begin{equation}
S(\omega)=\mathcal{F} \left[ \frac{\langle \mathcal{I}(t)\mathcal{I}(t+\tau) \rangle_{\rm{ss}}-\mathcal{I}^2_{\rm{ss}}}{\langle \mathcal{I}(t)\mathcal{I}(t) \rangle_{\rm{ss}} - \mathcal{I}^2_{\rm{ss}}} \right] = \mathcal{F}[\langle \sigma_z (t) \rangle], 
\end{equation}
where $\mathcal{F}[\langle \sigma_z (t) \rangle]$ is the Fourier transform of the (ensemble averaged) expectation value of the $\sigma_z$ operator.  From this response spectrum, we extract the Hamiltonian and decoherence parameters directly~\cite{Cole:05,Cole:06} for each spatial location across the sample.  When performing this parameter estimation process, it is not necessary to make any assumptions about the source or mechanisms producing the decoherence.  Simply mapping the induced decoherence as a function of position provides a dynamical image of the sample fluctuations.

In the limit of small $\kappa$, the model of Eq.~(\ref{eq:POVM}) is equivalent\cite{Brun:02} to more complicated master equation models\cite{Brun:02, Korotkov:01b, Zhang:05, Oxtoby:06, Oxtoby:08}.  Expanding the evolution to first order in both $\kappa$ and $\Delta t$, we can derive an equivalent Lindbladian master equation with an effective $\sigma_z$ decoherence channel of strength 
\begin{equation}\label{eq:MeasDecoh}
\Gamma_{\rm{meas}} = \frac{\kappa^2}{4 \Delta t},
\end{equation}
which corresponds to the measurement induced decoherence.  The measurement strength should be chosen such that this is smaller than both the sample induced decoherence and other intrinsic decoherence sources\footnote{This means operating well above the Korotkov-Averin bound\cite{Korotkov:01,Korotkov:01b} so that the sample decoherence effects dominate the signal.}.  The intrinsic noise sources include those introduced by the tip supporting the qubit, although ideally this decoherence is equal to or less than that introduced by a bulk substrate, owing to the geometry of the tip.

We can also calculate the information extracted at each measurement step, $\Delta t$, by looking at the reduction in entropy, $\Delta S_{\rm{E}}$, of the system.  Expanding for small $\kappa$, this gives
\begin{equation}\label{eq:Entropy}
\Delta S_{\rm{E}} = S_{\rm{E}}(\rho) - S_{\rm{E}}(\rho') = \frac{\kappa^2}{\log_e(4)} + \mathcal{O}(\kappa^3) 
\end{equation}
as the information obtained (in bits) from a single measurement of an initially mixed state $\rho = (\ket{0} \bra{0} + \ket{1} \bra{1})/2$.  As we increase $\kappa$ or the bandwidth, the amount of information obtained in a given time interval is increased, Eq.~(\ref{eq:Entropy}), at the expense of greater measurement induced decoherence, Eq.~(\ref{eq:MeasDecoh}).

The spatial resolution of the probe system is ultimately governed by the effective strength of the environmental decoherence as a function of distance.  Most decoherence channels (for solid-state qubits) depend, in some way, on the inverse of the separation between qubit and environment. 
The coupling between a single decoherence (point) source locate at $x_s$ and the probe qubit is, in general, proportional to $1/r^n$, hence the response of the qubit as a function of ($x$-)position across the sample is given by a exponentiated Lorenztian, 
\begin{equation}
\mathcal{R}(x) = \left[ \frac{1}{h_p^2+(x-x_s)^2} \right]^{n/2}.
\end{equation}
The Full Width at Half Maximum (FWHM) of this function gives the spatial resolution $\Delta x = 2 h_p \sqrt{2^{2/n}-1}$, where $h_p$ is the height of the probe above the sample.  By inspection, we see an electric dipole induced potential ($1/r^2$) has a FWHM $\Delta x = 2h_p$ whereas a magnetic dipolar interaction ($1/r^3$) has slightly better resolution, $\Delta x \approx 1.53 h_p$.

We now consider two examples in which we simulate the effects of a decohering environment on an example probe.  These examples provide both a straightforward illustration of the power of imaging the sample induced decoherence and a test of the feasibility using current and near future technology.

%%%%%%%%%%%%%%%%%%%%%%%%%%%%%%%%%%%%%%%%%%%%%%%%%%%%%%%%%%%%%%%%%%%%

\section{Example I: Imaging the distribution of background charge fluctuators}  
Our first example system comprises an electrostatic qubit interacting with a sample containing background charge fluctuators.  This example is of particular interest as background charge fluctuations have been the subject of intense scrutiny due to their relevance to solid-state quantum devices~\cite{Schriefl:06,Galperin:06}.  In this case, we can probe the spatial and frequency distribution of these fluctuators in a way which is not possible using current microscopy techniques.

A suitable probe qubit would be any of the myriad of charge based qubit designs such as quantum-dots\cite{Loss:98}, donors\cite{Hollenberg:04} or cooper-pair-boxs~\cite{Makhlin:01,You:07,Koch:07}.  A probe based on a cooper-pair-box (CPB) system provides a particularly good example as CPB qubits are now regularly produced experimentally\cite{Nakamura:99,Yamamoto:03,Majer:07} and the bias point of the system can be varied, resulting in a change in the sensitivity to different components of the environmental decoherence\cite{Schriefl:06}.  For generality, we model the qubit purely as a two-state system which interacts via a field dependent component in its Hamiltonian (see appendix~\ref{sec:MethodsII}) and therefore all the dimensions in this example are given in terms of a normalized length scale ($L$).  

%%%%%%%%%%
\begin{figure} [tb!]
\centering{\includegraphics[width=12cm]{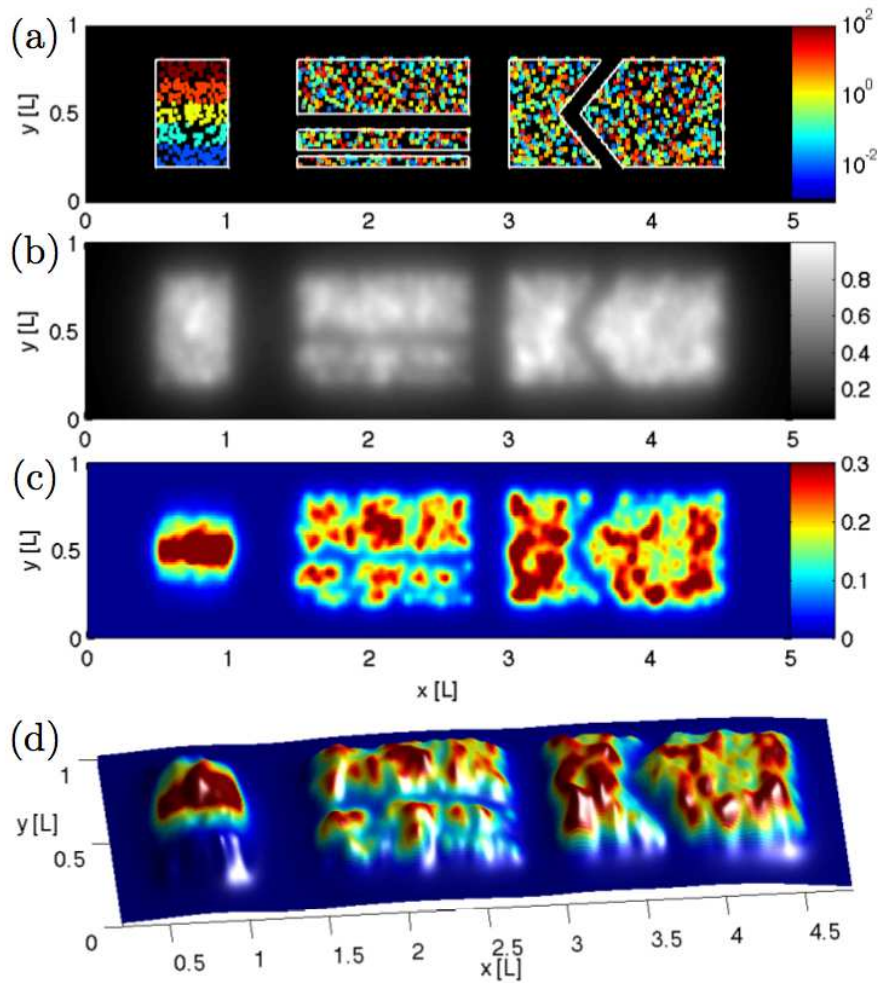}} 
\caption{(a) A fictitious sample with uniform distribution of fluctuators (coloured dots) within non-trivial spatial regions (outlined in white).  The fluctuators are modeled as point sources but have been enlarged here for clarity.  As we have not chosen a particular qubit architecture, all distances are in normalized length units ($L$) and the colour coding of the fluctuator locations depends on the frequency of the fluctuator, in units of the qubit probe frequency.  The region on the left is a calibration region where the fluctuator frequency varies with position in the region.  The other regions contain fluctuators with a uniform spatial and $1/f$ frequency distribution.  (b) Simulated response of the probe Hamiltonian interacting with a uniformly distributed bath of charge dipoles, with $h_p=0.05 L$.  The normalised variation of the Hamiltonian component ($\Delta$) induced by the background charges is plotted as the colour scale. (c) The effective decoherence (normalized) rate felt by the qubit as a function of position.  The qubit is now sensitive to the $1/f$ nature of the fluctuator bath, with a relatively small number of fluctuators contributed a large fraction of the decohering effect.   (d) Plots (b) and (c) have been combined using the value of (b) for the height and (c) for the colour.  The frequency selective nature of the image is apparent, where fluctuators closest to the probe in energy contribute the most decoherence signal (red peaks) where other peaks (blue) are to due to fluctuators which do not strongly decohere the qubit.\label{fig:HandGamma}}
\end{figure}%%%%%%%%%%

Let us consider a 2D surface which contains background charge fluctuators that we wish to image.  We use a simplified model of these fluctuators (see appendix~\ref{sec:MethodsII}) in the fast fluctuator limit, to illustrate the concepts.  In example II, we will consider a more specific example in the slow fluctuator limit.  We take the potential felt by the qubit as a simply electrostatic potential due to a charge defect dipole and simulate the effective \emph{decoherence field} felt by the qubit due to the fluctuator bath (defects in the sample).  In Fig.~\ref{fig:HandGamma}(a), we have an fictitious sample comprised of regions (outlined) containing a uniform fluctuator distribution with an area density of 1000 defects per square (all distances are in normalised units).  The frequencies of these fluctuators are then assumed to be distributed with a $1/f$ distribution.  We have also included a `calibration' region on the left of the sample, where the frequency of fluctuators is varied in a controlled fashion.  This will illustrate the frequency selective nature of the measurement process.

The variation of the probe Hamiltonian, $H_{xy}$, due to the presence of background charges is shown in Fig.~\ref{fig:HandGamma}(b).  The background charges are taken to be charge dipoles~\cite{Paladino:02, Galperin:06} coupling to $\Delta$, the $\sigma_z$ component of the qubit, as $1/r^2$ (see appendix~\ref{sec:MethodsII}) and the probe is positioned $h_p=0.05 L$ above the sample.  Initially assuming the fluctuators to be static, the spatial variation in the Hamiltonian coupling term is then a measure of the electric potential induced by the dipoles (a simple electrometer).  The state of the fluctuators will not be static (in general) and the total induced field will result from an average over the fluctuator states.  In Fig.~\ref{fig:HandGamma}(c), we calculate the decoherence felt by the probe due to the combined effects of all the fluctuators in our fictitious sample.  The resolution \emph{for this example} increases by a factor of $1.55$, compared to measuring the Hamiltonian, as $\Gamma \propto S(\omega) \propto 1/r^4$.  Comparing these images, we see that the decoherence measurement is more sensitive to the fluctuators whose frequency are closest to the probe energy.  It is known that, for a $1/f$ bath, a relatively small fraction of the total fluctuators contributed a large amount of the total decoherence~\cite{Schriefl:06} but here we see it strikingly depicted in the images.

To further demonstrate the utility of mapping both Hamiltonian and decoherence components simultaneously, Fig.~\ref{fig:HandGamma}(d) is a combined plot of both  Fig.~\ref{fig:HandGamma}(b) and Fig.~\ref{fig:HandGamma}(c).  In this case, the effective field (static) is a surface plot with colouration corresponding to the decoherence rate (dynamics).  The vertical scale of this `mountain range' plot corresponds to the varying electric potential across the sample.  The red `mountains' are fluctuators with frequency close to the qubit transition frequency while blue corresponds to far off resonant fluctuators.  In principle, the frequency sensitivity of the qubit can be tuned to probe different components of the decoherence, to obtain more information about the spectral characteristics ($1/f$, Ohmic etc.\ ) as a function of position.

%%%%%%%%%%
%\begin{figure} [tb!]
%\centering{\includegraphics[width=8cm]{example_3hs.eps}} 
%\caption{Simulated response for a probe interacting with a uniformly distributed bath of $1/f$ fluctuators for several different probe heights above the sample, h = 0.01, 0.05, 0.1 $L$.  The (normalised) decoherence rate induced by the background charges (as calculated using Eq.~\ref{eq:GRGamma}) is plotted as the (truncated) colour scale.  The individual fluctuators are only resolved when the probe height is commensurate with the mean fluctuator separation, in this case approximately $0.3 L$.\label{fig:eg1_3hs}}
%\end{figure}%%%%%%%%%%

%%%%%%%%%%%%%%%%%%%%%%%%%%%%%%%%%%%%%%%%%%%%%%%%%%%%%%%%%%%%%%%%%%%%%%%%%%%%%%%

\section{Example II: Imaging the position and spin occupation of Ferritin molecules}  
Now, we choose to use both a probe qubit and sample which have been well studied experimentally.  In contrast to the previous example, where all parameters were dimensionless, here we investigate a specific implementation to evaluate the technical feasibility of the proposal.  We consider the imaging of (bio)molecules  with large uncompensated spin, such as Horse-spleen Ferritin\cite{Awschalom:92,Tejada:97} or $\rm{Fe}_8$\cite{Barco:00}.  Here, the point is neither to image an individual spin\cite{Rugar:04} or image the location of the molecules\cite{Hansma:94, Birdi:03, Pakes:04} as both can be done with existing technology.  We show that qubit probe imaging can both map the location of the spins and probe their magnetic dynamics.

\begin{figure} [tb!]
\centering{\includegraphics[width=12cm]{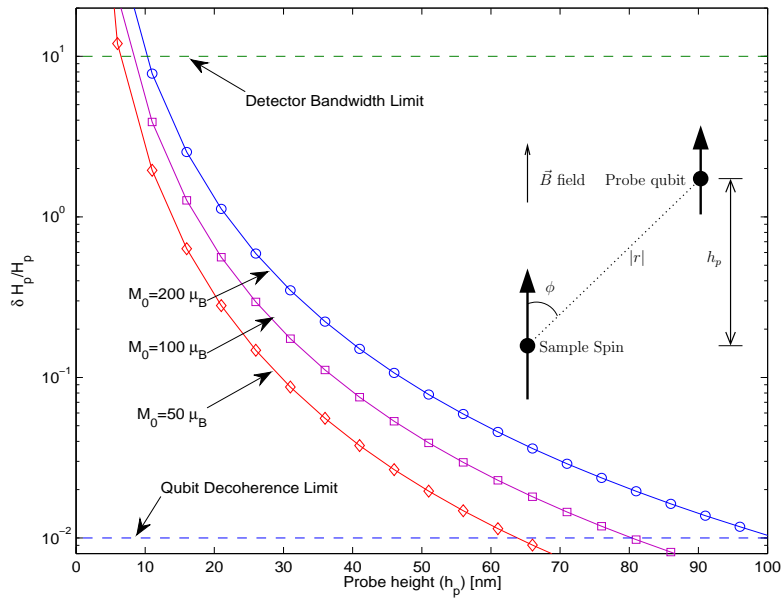}} 
\caption{Variation in strength of the maximum signal ($\phi=0$) as a function of probe height for different sized spins.  The signal is the fractional variation in the probe spin Hamiltonian due to the influence of the sample spin.  The upper and lower threshold limits come from the measurement bandwidth and decoherence respectively. The insert illustrates the simplified magnetic dipolar coupling model, which depends on the separation, orientation and magnetic moment of both the probe and sample spins.\label{fig:dipolardiag}}
\end{figure}

The decoherence introduced at the probe qubit will be a function of both the interaction strength and the flipping rate of the sample spin(s).  As we are considering large sample spins in a static magnetic field at low temperature, we will assume that the flipping rate is slow on the scale of the probe Hamiltonian.  The spectral response of the qubit is split, with the separation between the peaks giving the effective difference in the Hamiltonian between the two sample spin states (see appendix~\ref{sec:MethodsIII}).  We compute the coupling strength between the spins, given the simplified magnetic dipolar model, as illustrated in the insert of Fig.~\ref{fig:dipolardiag}.  The ability to resolve the induced coupling is ultimately limited by the total effective decoherence rate of the qubit.  The ratio of the peaks also gives a measure of the relative spin populations, which in turn relates to the effective temperature and/or flipping mechanism.

We use a nitrogen-vacancy (NV) centre in diamond as our probe spin, as this has been shown to be a controllable, well isolate spin system which displays stable quantum coherent properties up to room temperature~\cite{Jelezko:01,Jelezko:04a,Jelezko:04b,Charnock:01,Kennedy:02,Gaebel:06,Hanson:06,Hanson:08} and has even been used as a nanoscale scanning magnetometer~\cite{Balasubramanian:08, Maze:08}, in a similar cantilever configuration to that which we envisage.  We then couple this probe to a bulk spin of order $M_0=50-200\mu_B$ and include the effects of intrinsic decoherence and finite measurement bandwidth.  Using experimentally realistic parameters for both the probe qubit and the sample spin, we plot the response of the system as it passes over the spin.  An alternative probe qubit system would be to use a microSQUID or flux qubit~\cite{Ilichev:07, Shnyrkov:07, Makhlin:01}. 

For our example, we use known system parameters for an NV centre driven by a microwave loop and readout via a laser probe measurement~\cite{Wrachtrup:06}.  We take the Rabi frequency of the qubit to be $10 \rm{MHz}$ and the measurement bandwidth $\rm{BW} = 100 \rm{MHz}$ but not necessarily `strong'.  The intrinsic decoherence rate is approximately $100$ times slower than the Rabi frequency.  The measurement strength $\kappa$ is chosen such that the measurement induced decoherence is weaker than the intrinsic decoherence, for a given detector bandwidth.

\begin{figure} [tb!]
\centering{\includegraphics[width=12cm]{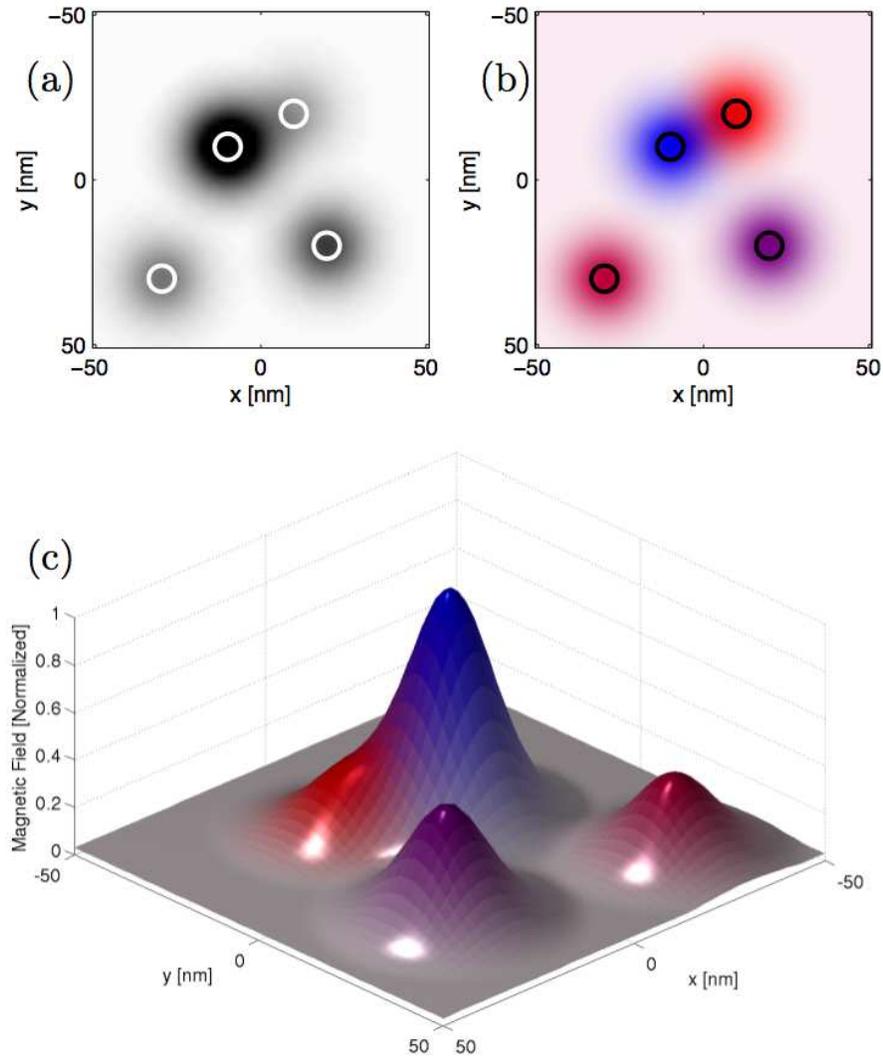}}  
\caption{(a) Magnetic field strength as measured via the variation in the probe qubit Hamiltonian. The pixel intensity represents the normalized fractional variation in the probe spin Hamiltonian.  The detection limits discussed earlier for maximum and minimum detectable field are included in this calculation.  (b) The effective temperature of each spin, based on the population of the ground and excited states.  The intensity in this plot is given by the effective field strength of each spin signal separately.  The white/black circles indicate the position and diameter ($8 \rm{nm}$) of the sample spin and enclosing molecules. (c) Surface plot of the magnetisation with colouration based on the fraction of time spent in the excited spin state.  The height and colourscale are normalised as per (a) and (b) respectively.  The grey colouration corresponds to a region where the peak splitting is not large enough to resolve the spin populations with these measurement parameters.\label{fig:image2d}}
\end{figure}

Fig.~\ref{fig:dipolardiag} shows the maximum fractional Hamiltonian component as a function of probe height for three different spin samples.  The measurement window is defined between the decoherence rate and the measurement bandwidth.  The probe height can vary over almost 100 nm and still provide a detectable signal.
%%%%%%%%%%
%%%%%%%%%%
In Fig.~\ref{fig:image2d}, we create an image from a fictitious sample consisting of 4 mesoscopic spins with varying net magnetisation.  We assume the spins are in thermal equilibrium with the sample substrate (which we have set at $T = 4K$) and that they are flipping due to thermal processes\footnote{The exact details of the flipping process are unimportant for this example.  We assume they are dominated by thermal processes and obey Boltzmann statistics.}.  The magnetisations are $M_0 = 50$, $70$, $100$ and $200 \mu_B$ and the average population of the excited state is given by a Boltzmann distribution for a background magnetic field of $B = 0.1 \rm{T}$ and temperature $T = 4\rm{K}$.  The spatial resolution of the probe position is a $50\times50$ grid, giving 2500 points over $10000 \rm{nm}^2$ and the probe height was set to $h_p = 20 \rm{nm}$.  Fig.~\ref{fig:image2d}(a) shows the measured magnetic field over the sample.  Note that the probe in this mode (purely acting as a magnetometer) does not successfully resolve two of the spins.

%%%%%%%%%%
%%%%%%%%%%

Fig.~\ref{fig:image2d}(b) shows the measured decoherence field over the same sample.  As each spin has a different magnetisation, the decoherence effects (in this case splitting of the spectral response peak) resulting from each spin are different.  The ratio of the two split peaks provides the population of the spin states, which is directly related to the magnetisation and effective temperature of the sample spin.  In this plot, the ratio of the split peaks has been used to colour code the data, with blue indicating a large spin magnetisation (or low effective temperature) and red a small magnetisation (high temperature).  The intensity of the colour is given purely by the amount of signal available from each decoherence source (compared to the probe spin's intrinsic decoherence), whereas in Fig.~\ref{fig:image2d}(a), the intensity was proportional to the total induced field.  Finally, we can combine this data to produce a plot showing the field intensity with each decoherence source (mesoscopic spin) tagged based on its effective temperature.  This is shown in Fig.~\ref{fig:image2d}(c), where the existence of all four spins can be detected based on the colour tagging, in contrast to the magnetometer scan alone.

%%%%%%%%%%%%%%%%%%%%%%%%%%%%%%%%%%%%%%%%%%%%%%%%%%

\section{Image acquisition rate and noise}
\begin{figure} [tb!]
\centering{\includegraphics[width=12cm]{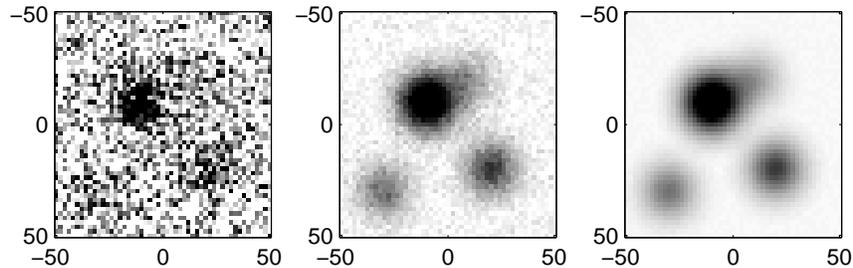}} 
\caption{The image from Fig.~\ref{fig:image2d}(a) with simulated noise stemming from a finite measurement time.  The images from left to right simulate dwell times of $t_{\rm{dwell}}=2\mu \rm{s}$, $200 \mu \rm{s}$ and $20 \rm{ms}$ which give a total acquisition time for a $50\times50$ grid of $t_a = 5 \rm{ms}$, $0.5 \rm{s}$ and $50 \rm{s}$ respectively.  The average pixel intensity variance is $10^{-1}$, $10^{-3}$ and $10^{-5}$ for the three images.\label{fig:uncertexample_m2}}
\end{figure}

While we have demonstrated that new information can be obtained by looking at the induced decoherence, this is only useful if the information can be obtained within an experimentally accessible time.  Using the measurement model discussed earlier, we can estimate the parameter uncertainties in the Hamiltonian characterisation process~\cite{Cole:05,Cole:06}.  Retaining the parameters from Example II, we calculate the noise expected for a finite dwell time ($t_{\rm{dwell}}$) on each pixel and the total image acquisition time ($t_a$).  Fig.~\ref{fig:uncertexample_m2} shows Fig.~\ref{fig:image2d}(a) with the noise resulting from a finite bandwidth and dwell time.  The image acquisition times (for 2500 pixels) varies between $5 \rm{ms}$ and $50 \rm{s}$ for pixel error variances which range from $10^{-1}$ to $10^{-5}$.  

To allow fast image acquisition times (using continuous a measurement mode), it is important to have both large detector bandwidth and a large ratio of bandwidth to intrinsic decoherence rate $\rm{BW}/\Gamma_q$.  The qubit transition frequency is less important, provided it is at least an order of magnitude greater than the intrinsic decoherence.  However, a tunable qubit frequency is advantageous as many types of decoherence depend strongly on the frequency at which they are probed.

To estimate detector noise as a measure (rather than acquisition time), we take recent noise limit estimates for magnetometry based on NV centres~\cite{Taylor:08, Degen:08} as a guide.  In this case, the measurement is a `DC' measurement ($\rm{BW}=10 \rm{kHz}$) with a sensitivity of $20 \rm{nT}/\sqrt{\rm{Hz}}$~\cite{Taylor:08} and a $10\rm{MHz}$ qubit frequency.  This results in an additional contribution to the pixel variance of $\approx 10^{-5}$ but still results in a clear image (see Fig.~\ref{fig:uncertexample_m2}).
 %%%%%%%%%%
%%%%%%%%%%

\section{Conclusions}
We have proposed and theoretically investigated a fundamentally new and complementary imaging mode which takes advantage of current qubit technology and modern techniques for characterising few-state quantum systems.  Mapping the induced decoherence across a sample indirectly images the dynamics of the environment, providing a new window into the microscopic world with wide applications to spin and charge fluctuations and transport in both solid-state and biological systems.  In contrast to a scanning qubit magnetometer, the decoherence signature is sensitive to fluctuations even when the time-averaged field is close to zero.  Such processes abound in nature, in which the time dependent behaviour of electromagnetic fields stems directly from dynamical processes of interest, such as ion flow, spin fluctuations or conformational change.   The device may be realised using a range of architectures, each of which will be sensitive to a particular electromagnetic effect within a sample.  While we have discussed several specific examples using near future technology, the basic requirements for a decoherence microscopy already exist in the form of nano-scale positioning and individual two-state quantum systems which can be initialized, controlled and measured.  Hence, this technique has the potential to open a new window for imaging nanoscale processes in the physical and life sciences.

\section*{Acknowledgments}
The authors would like to thank N. Oxtoby, C. Hill, G. Milburn and G. Sch\"on for useful comments on the manuscript.  This work was supported by the Australian Research Council.  JHC acknowledges the support of the Alexander von Humboldt Foundation and LCLH is the recipient of an Australian Research Council Australian Professorial Fellowship (DP0770715).

\appendix

\section{Example I - Decoherence due to a bath of $1/f$ fluctuators}\label{sec:MethodsII}
We wish to use a simple model of the decoherence felt by a probe (charge-)qubit interacting with a bath of $1/f$ fluctuators~\cite{Paladino:02,Galperin:06,Schriefl:06}. The numerical parameters used for such a model vary greatly depending on the system and even from sample to sample.  Here we are interesting in the functional dependence, rather than the explicit values.  We draw heavily from references~[\onlinecite{Paladino:02}] and~[\onlinecite{Galperin:06}] as an example.

Consider a qubit interacting with a bath of fluctuators via an interaction Hamiltonian $H_{\rm{int}}$ which has the form
\begin{equation}
 H_{\rm{int}} = \sigma_z \sum_j v_j b_j^\dag b_j
\end{equation}
where $\sigma_z$ acts on the qubit and $b_j$ ($b_j^\dag$) destroys (creates) an electron in a localized state in the bath.

The spectral response from the $j$th fluctuator (in the fast fluctuator limit) is given by
\begin{equation}
s_j(\omega)\propto \frac{v_j^2  \gamma_j}{\gamma_j^2+\omega^2}
\end{equation}
where $v_j$ is the strength of the fluctuator which fluctuates with rate $\gamma_j$.  The total spectral response is then a sum over each of the fluctuators, 
\begin{equation}
S(\omega) = \sum_j s_j(\omega)
\end{equation}
and a simplified Golden rule model gives the relaxation rate ($\Gamma_-^{GR}$) and dephasing rate ($\Gamma_2^{GR}$) as
\begin{equation}\label{eq:GRGamma}
\Gamma_2^{GR}=\frac{1}{2}\Gamma_-^{GR}=\frac{1}{4}S(E_J)
\end{equation}
where $E_J$ is the tunnelling energy of the qubit.
Given a functional form for how $v_j(r)$ depends on the fluctuator/qubit separation $r$, we can then compute the effective decoherence rate felt by the qubit.

\section{Example II - Magnetic dipolar coupling between a probe and sample spin}\label{sec:MethodsIII}
For this analysis, we will use a simplified (but quite general model) consisting of a probe spin interacting via the magnetic dipolar interaction with a much larger sample spin (see insert of Fig.~\ref{fig:dipolardiag}).  The sample spin can be considered as a Ferritin, $\rm{Fe}_8$ or other mesoscopic molecule with a net magnetic moment $M_0\gg1 \mu_B$.  As the spin is large (and to simplify the analysis of decoherence) we will assume that it is in thermal equilibrium with the sample environment and therefore the average magnetisation and spin flip rates are given by the standard thermodynamic quantities.  In making this assumption, we ignore quantum mechanical effects between the probe and sample spins and treat the effect of the sample spin on the probe as a fluctuating classical field.

The magnetic dipolar interaction is given by~\cite{Poole:72}
\begin{equation}
H_{\rm{dip}} = \left( \frac{\mu_0}{4 \pi}\right) \hbar^2 \gamma_p \gamma_s \left[ \frac{\vec{P}.\vec{S}}{r^3} - \frac{3 (\vec{P}.\vec{r})(\vec{S}.\vec{r})}{r^5}\right]
\end{equation}
where $\gamma_p$ and $\gamma_s$ are the probe and sample spin gyromagnetic ratios, $\vec{r}$ is the vector separation between the spins and $\vec{P}$ and $\vec{S}$ are the probe and sample spin operators.

We assume the system is bathed in a global magnetic field $B$ which orientates both the sample and probe spins\footnote{We ignore the effects of crystal field terms on both the orientation of the sample and probe spin, though this does not decrease the generality of the result.} and sets their energy scales.  The dipolar interaction is then given by the separation between the spins $r$ and the angle subtended between the spin orientation and the vector separating the spins $\phi$, as illustrated in the insert to Fig.~\ref{fig:dipolardiag}.  The simplified coupling Hamiltonian is therefore
\begin{equation}
H_{\rm{dip}}=\left(\frac{\mu_0}{4 \pi}\right)\left(\frac{\hbar^2 \gamma_p \gamma_s}{r^3}\right)(3 \cos^2 \phi -1) P_z.S_z,
\end{equation}
which is a purely Ising type interaction whose strength depends on both the separation and angle between the spins.
The perturbing effect of this dipolar Hamiltonian can then be measured in the spectral response of the qubit, giving a direct link to both the spin state and magnetisation of the sample spin.

\section*{References}
%\bibliographystyle{iopart-num}
%\bibliography{decohprobe}

\end{document}